\documentclass[aps,prl,showpacs,twocolumn]{revtex4}
\usepackage{graphicx}

\begin{document}

\title{Polarization tomography of metallic nanohole arrays}
\author{E. Altewischer, C. Genet, M.P. van Exter, J.P. Woerdman}
\affiliation
{Huygens Laboratory, Leiden University, P.O. Box 9504, 2300 RA
Leiden, The Netherlands}
\author{ P.F.A. Alkemade}
\affiliation
{Faculty of Applied Science, Delft University of Technology,
Rotterdamseweg 137, 2628 AL Delft, The Netherlands}
\author{A. van Zuuk, E.W.J.M. van der Drift}
\affiliation
{DIMES, Delft University of Technology, PO Box 5053, 2600 GB
Delft, The Netherlands}
\date{\today}

\begin{abstract}
We report polarization tomography experiments on metallic nanohole
arrays with square and hexagonal symmetry. As a main result, we
find that a fully polarized input beam is partly depolarized after
transmission through a nanohole array. This loss of polarization
coherence is found to be anisotropic, i.e. it depends on the
polarization state of the input beam. The depolarization is
ascribed to a combination of two factors: i) the nonlocal response
of the array due to surface plasmon propagation, ii) the non-plane
wave nature of a practical input beam.
\end{abstract}
\maketitle

Currently, there is much interest in the optical properties of
thin metal films perforated with arrays of subwavelength holes, or
nanohole arrays. The optical transmission of these arrays shows a
strongly peaked spectrum with anomalously large transmission peak
values; this is usually ascribed to resonant excitation of
propagating surface electromagnetic waves or surface plasmons
(SPs)\cite{Ebbesen,Ghaemi}, which exist due to the fact that the
real part of the dielectric function of a metal is smaller than -1
for frequencies below the plasma frequency\cite{Raether}. There is
no restriction on the imaginary part, however. Indeed, similar
anomalous transmission spectra have recently been observed in
scaled versions of a metal hole array in the (sub)millimeter-wave
domain, where the dielectric function of a metal is dominantly
imaginary, but long-range SPs still
exist\cite{Beruete,Gomez,Miyamaru}. This suggests that the
propagating nature of the SP is the key to understanding the
anomalous transmission. In this Letter we focus on the
polarization properties of the anomalous transmission, and show
that these, too, are strongly influenced by the propagating nature
of the SPs .

So far, polarization properties of nanohole arrays have been
studied in a limited context: a beam with a given uniform state of
polarization (SOP$_{in}$) is transformed, by an anisotropic array,
in a different uniform state of output polarization (SOP$_{out}$)
\cite{Elliott,Gordon}. This corresponds to a unitary mapping on
the Poincar\'{e} sphere; for instance, a rectangular array or a
square array with elliptical holes acts as a birefringent and/or
dichroic element, which may convert a linear SOP into an
elliptical SOP, conserving polarization coherence. This is not
what we deal with in the present Letter; instead we focus on cases
where the \textit{degree} of polarization (DOP) is reduced,
$\mathrm{DOP}_{out}<\mathrm{DOP}_{in}$\cite{Azzam,LeRoy}. To
underline this point, we have chosen for our experiments square
and hexagonal arrays, i.e. arrays that, for symmetry
reasons\cite{Genet}, cannot modify the SOP for plane-wave
illumination at normal incidence. As we will show,
\textit{de}polarization occurs when two (quite common) conditions
are fulfilled: (i) the response of the array is nonlocal due to SP
propagation, and (ii) the input beam is \textit{not} a plane wave
(but e.g. a Gaussian beam, with a finite numerical aperture (NA)).

In general, depolarization occurs when an optical system acts
non-uniformly on polarization within the (spatial or temporal)
bandwidth of the incident wave, thereby coupling polarization to
other degrees of freedom. The two most widely used formalisms to
describe the polarization properties of a linear optical system
are due to Jones and Mueller. In both cases, the input and output
SOPs are represented by column vectors, which are connected by
either a $2 \times 2$ Jones matrix or a $4\times 4$ Mueller
matrix\cite{Azzam,LeRoy}, describing the action of the optical
system. The Jones formalism is applicable only to situations in
which the light is temporally coherent and spatially-uniform
polarized. The Mueller formalism deals with the Stokes parameters,
which represent (spatial or time) averages of the polarization
properties of the light, and as such, is also capable of handling
partially polarized and incoherent waves. Experimentally, a study
of depolarization requires therefore a measurement of the Mueller
matrix by a tomographic method\cite{LeRoy}. We report here such
polarization tomography experiments on nanohole arrays and
interpret the results in the context of SP propagation.

We start by recapitulating the essence of our theoretical model
\cite{Genet}. The input and output optical fields of the array are
related via a non-local linear response as
$\vec{E}_{out}(\vec{r},\omega)=\int
t(\vec{r}-\vec{r^\prime},\omega)\vec{E}_{in}(\vec{r^\prime},\omega)\,d\vec{r^\prime}$.
In the far-field, or Fourier domain, this is equivalent to
$\vec{E}_{out}(\vec{k_t},\omega)=t(\vec{k_t},\omega)\vec{E}_{in}(\vec{k_t},\omega)$,
where $\vec{k_t}$ is the transverse wavevector component; the
output $\vec{k_t}$ is equal to the input $\vec{k_t}$ for the
zeroth-order transmission. If the non-local response depends on
polarization, the elements of the $2\times2$ transmission tensor
$t(\vec{k_t},\omega)$ exhibit a different angular dependence and
the output field $\vec{E}_{out}(\vec{k_t},\omega)$ can have a
spatially-dependent polarization even for a polarization-pure
input field. This necessitates a Stokes vector description, which
considers only spatially-integrated intensities. The four
components of the Stokes vector are
$S_0=I_{0^\circ}+I_{90^\circ}=I_{45^\circ}+I_{135^\circ}=I_{\sigma+}+I_{\sigma-}$,
and $S_1=I_{0^\circ}-I_{90^\circ}$,
$S_2=I_{45^\circ}-I_{135^\circ}$ and
$S_3=I_{\sigma+}-I_{\sigma-}$, with $I_i$ the intensity of the
polarization component denoted by the subscript $i$\cite{LeRoy}.
After spatial integration, the transmission process can be
captured in a simple relation $S_{out}=MS_{in}$, which relates the
input and output Stokes vectors through the $4\times4$ Mueller
matrix $M$. For ideal square and hexagonal arrays the Mueller
matrix has been predicted to be diagonal (no mixing of Stokes
parameters)\cite{Genet}. For hexagonal arrays, the additional
symmetry relation $M_{11}=M_{22}$ holds.

The magnitudes of the diagonal elements $M_{00}$, $M_{11}$,
$M_{22}$ and $M_{33}$ depend on the product of the SP propagation
length and the wave vector spread of the input beam, $\ell_{SP}
\Delta k_t$. A full theoretical description thereof would require
a microscopic model; however, from physical considerations it can
be seen that an appreciable deviation of $M_{ii}/M_{00}$ from 1
requires $\ell_{SP} \Delta k_t \gtrsim 1$. In any case, there will
be no depolarization if either there are no propagating waves
($\ell_{SP} =0$) or there is plane-wave illumination ($\Delta
k_t=0$); \textit{both} propagation and wave vector spread are
necessary.

\begin{figure}
\centerline{\includegraphics{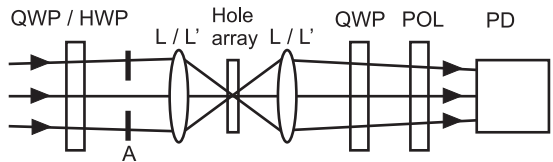}} \caption{Experimental
setup to measure the Mueller matrix of nanohole arrays. The
incident polarization state is set by a quarter-wave plate (QWP)
or half-wave plate (HWP), the output polarization state is
measured with a combination of a QWP, polarizer (POL) and
photodiode (PD). Lenses L or L' form a symmetric telescope, at the
focus of which the hole array is placed; a variable aperture A
sets the numerical aperture of the illumination.}
\end{figure}

Figure 1 shows the experimental setup used to measure the Mueller
matrices. A linearly polarized Titanium-Sapphire laser at a
wavelength of 810~nm, being the approximate resonance wavelength
of the hole arrays (cf. Fig.~2), illuminates the input lens ($L$
or $L^\prime$) of a symmetric telescope. After transmission
through a hole array, positioned at the focus of the telescope,
the light is imaged onto a photodiode. The SOP of the incident
light is set by a rotatable quarter-wave or half-wave plate in
front of the first lens. The Stokes parameters at the output are
measured with a rotatable quarter-wave plate and polarizer
positioned in front of the photodiode.

We have characterized the arrays with two types of illumination.
Almost plane-wave illumination, with a numerical aperture of
NA~=~0.01 ($\Delta k_t \approx 0.08 \mu\mathrm{m}^{-1}$), was
provided by focussing the 1~mm diameter laser beam with a lens $L$
of f~=~50~mm. Focussed illumination (up to NA~$\approx 0.15$ or
$\Delta k_t \approx 1.2 \mu\mathrm{m}^{-1}$) was obtained by
focussing the laser beam on a 10~$\mu$m diameter mode-cleaning
pinhole (not shown) to homogeneously illuminate a lens $L^\prime$
(f=15~mm at 40~cm from the pinhole) through a variable aperture
$A$. The polarization isotropy of all optical components was
checked by measuring the Mueller matrices of both setups in the
absence of hole arrays. These matrices were practically equal to
the identity matrix, with individual elements deviating by not
more than 0.02 (typically 0.008).

\begin{figure}
\centerline{\includegraphics[width=7cm]{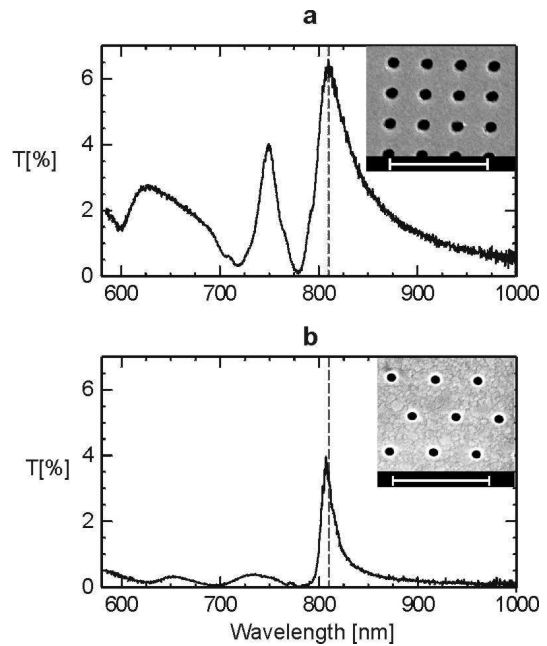}}
\caption{Transmission spectra of nanohole arrays under almost
plane-wave illumination at normal incidence for the a) square and
b) hexagonal array. The dashed vertical line indicates the
resonance wavelength of 810~nm used in the experiments. The insets
show scanning electron beam micrographs of the arrays; the scale
bars correspond to 2~$\mu$m.}
\end{figure}

Our arrays were fabricated in Au films on glass substrates. We
used a square array made with electron-beam lithography, identical
to the one used in Ref.~\cite{Altewischer}, with a lattice spacing
of 700~nm and a nominal hole diameter of 200~nm, and a hexagonal
array made with ion-beam milling, with a lattice spacing of 886~nm
and a nominal hole diameter of 200~nm. SEM pictures and
transmission spectra of both arrays under almost plane-wave
illumination at normal incidence are shown in Fig.~2. Both arrays
show a resonance wavelength of 810~nm, which is marked by a dashed
vertical line; the polarization experiments were performed at this
wavelength. The resonances at 810~nm correspond to SPs propagating
in the $(\pm1,\pm1)$ direction at the metal-glass interface for
the square array, and the (six-fold degenerate) $(1,0,0)$
direction at the metal-air interface for the hexagonal array (the
labeling is with respect to the reciprocal lattice vectors). The
linewidths are 40~nm and 25~nm, for square and hexagonal arrays,
respectively, from which we estimate $\ell_{SP} \approx 2\mu$m and
$\ell_{SP} \approx 4\mu$m, respectively\cite{Altewischer2}.

\begin{figure*}
{\includegraphics{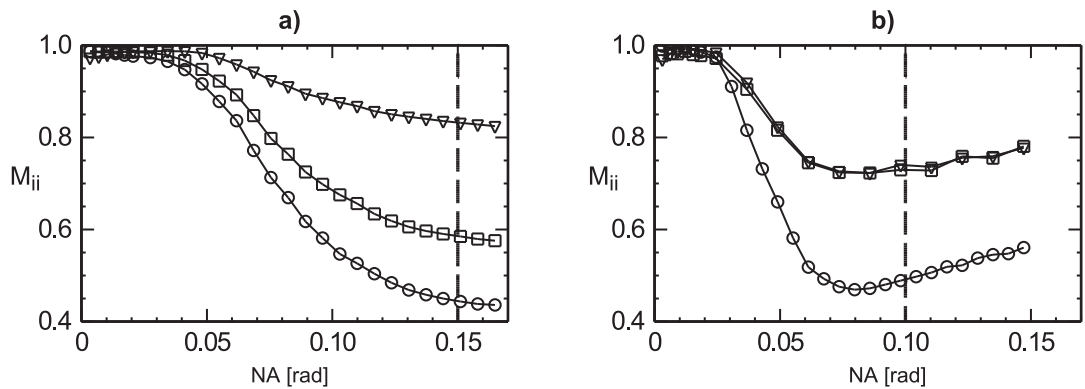}}
\caption{Diagonal elements $M_{ii}$ of the normalized Mueller
matrix as a function of the numerical aperture of the incident
light beam on the \textbf{a})~square and \textbf{b})~hexagonal
hole array. Squares indicate $M_{11}$, triangles $M_{22}$, and
circles $M_{33}$. Dashed vertical lines indicate the numerical
apertures at which the full Mueller matrices were measured (see
Table).}
\end{figure*}

Figure~3 shows the dependence of the diagonal elements of the
Mueller matrix on the NA of the incident light for both arrays
(the plotted elements $M_{11}$, $M_{22}$ and $M_{33}$ are
normalized with respect to $M_{00}$). The figure shows that, for
the case of almost plane-wave illumination
($\mathrm{NA}\approx0$), the $M_{ii}$ values are close to 1 for
both arrays. There is no depolarization, as $\Delta k_t \approx
0$. However, for increasing wavevector spread, or decreasing spot
size on the array, the depolarizing effect of the arrays quickly
increases. Furthermore, the depolarization is clearly anisotropic,
i.e. the amount of depolarization depends on the input SOP. For
the hexagonal array the depolarization sets in faster upon
increasing NA than for the square array, due to its smaller
resonance linewidth.

The effect of the different array symmetries on the curve shapes
are prominent. For the square array (Fig.~3a), the observed
inequality $M_{22}> M_{11}$ shows that there is less
depolarization for an input polarization along either of the array
diagonals than for a polarization along the main axes. This
observation is consistent with the $(\pm1,\pm1)$ propagation
directions of the resonantly excited SPs on the metal-glass
interface; as SPs are mainly longitudinally polarized, they
preserve polarization along their propagation direction. However,
the deviation of $M_{22}$ from 1 indicates that the $(\pm1,\pm1)$
SPs are not the only SPs involved in the transmission process;
other (non-resonant) SPs on both surfaces apparently
contribute\cite{Altewischer2}. For the hexagonal array (Fig.~3b)
the theoretical equality $M_{11}=M_{22}$ also holds quite well.
For both arrays, $M_{33}$ is the smallest diagonal element. This
shows that depolarization is most dramatic for circularly
polarized light, a case that has not been studied before. We note
that $M_{33}\approx M_{11}+M_{22}-1$ for both arrays, as follows
from an extension of the symmetry-based theory used here by
explicit modelling of SP propagation; this will be discussed
elsewhere\cite{Altewischer3}.

\begin{table}
\caption{Mueller matrices measured under four different
conditions. Normalization constants (absolute intensity
transmission for unpolarized light) of the matrices are 1.4\%
(hexagonal NA=0.01), 0.47\% (hexagonal NA=0.10), 7.1\% (square
NA=0.01) and 2.2\% (square NA=0.15). Each ($4\times 4$) matrix was
determined with 36 measurements for increased accuracy. Diagonal
elements are boxed for convenience, off-diagonal elements, which
indicate non-perfect array symmetry, are underlined.}
\begin{tabular}{l c c} \hline Measurement & & Mueller matrix M
\\ \hline
\\
$\begin{array}{l} \mathrm{Square~array} \\\mathrm{NA=0.01} \end{array}$& &$\left[ \begin{array}{rrrr} \fbox{\,1.00} &0.00 &-0.02 &0.00 \\
-0.02 &\fbox{\,1.01} &-0.01 &0.00\\
0.00 &0.00 &\fbox{\,1.01} &-0.01\\
0.00 &0.00 &0.02 &\fbox{\,0.99}
\end{array}\right]$ \vspace{5mm}\\

$\begin{array}{l} \mathrm{Square~array} \\\mathrm{NA=0.15} \end{array}$ & &$\left[ \begin{array}{rrrr} \fbox{\,1.00} &0.01 &-0.02 &0.01 \\
0.01 &\fbox{\,0.55} &-0.04 &-0.03\\
-0.04 &0.02 &\fbox{\,0.84} &0.01\\
-0.01 &0.01 &0.01 &\fbox{\,0.41}
\end{array}\right]$ \vspace{5mm}\\

$\begin{array}{l} \mathrm{Hexagonal~array} \\\mathrm{NA=0.01} \end{array}$& &$\left[ \begin{array}{rrrr} \fbox{\,1.00} &0.06 &\underline{-0.14} &0.00 \\
0.04 &\fbox{\,1.00} &0.00 &\underline{-0.07}\\
\underline{-0.12} &0.01 &\fbox{\,1.01} &-0.03\\
0.00 &\underline{0.06} &0.02 &\fbox{\,0.97}
\end{array}\right]$
\vspace{5mm}\\

$\begin{array}{l} \mathrm{Hexagonal~array} \\\mathrm{NA=0.10} \end{array}$& &$\left[ \begin{array}{rrrr} \fbox{\,\,1.00} &0.03 &\underline{-0.11} &0.00 \\
0.02 &\fbox{\,0.78} &0.00 &\underline{-0.08}\\
\underline{-0.13} &0.01 &\fbox{\,0.78} &0.00\\
0.00 &\underline{0.06} &0.02 &\fbox{\,0.51}
\end{array}\right]$

\end{tabular}
\end{table}

The full Mueller matrices for both arrays are shown in the Table.
They were measured with nearly plane-wave illumination (NA~=~0.01
or $\ell_{SP}\Delta k_t\approx0.2-0.3\ll 1$), and with
illumination at NA~=~0.15 ($\ell_{SP}\Delta k_t\approx2$) for the
square and NA~=~0.10 ($\ell_{SP}\Delta k_t\approx 3$) for the
hexagonal array (indicated by the dashed vertical lines in
Fig.~3). The diagonal elements (marked with boxes in the Table)
conform to the discussion given above. The off-diagonal elements
of a perfectly symmetric square or hexagonal array should
theoretically be zero. For our square array they are indeed
relatively small and do not show any systematic behavior. For the
hexagonal array however, these elements are much larger both for
plane-wave and focussed illumination; this array apparently does
not have perfect hexagonal symmetry. Furthermore, the off-diagonal
elements have a clear pattern and similar values in both cases,
where the ``odd'' off-diagonal elements $M_{02}$, $M_{20}$,
$M_{13}$ and $M_{31}$ (underlined in the Table) are substantially
larger than the others. This pattern was checked to be present
also for an intermediate NA of~0.03. The pattern is compatible
with a birefringent/dichroic $45^\circ$ axis, which is apparently
due to array errors, such as a spatially variant lattice spacing
or ellipticity of the holes. These errors could be created by
alignment errors or even intrinsic imperfections in the ion beam
optics (astigmatism and deflection errors).

From a general perspective, Mueller tomography can give new
insight into the physical mechanisms active in hole arrays. It
would be interesting to do Mueller tomography on metal hole arrays
in the (sub)millimeter-wave regime, as SPs propagate much farther
in this part of the spectrum, which is expected to increase the
depolarization. Another area of interest is the connection between
classical polarization properties and entanglement
degradation\cite{Altewischer,Altewischer3}. Our work also shows
that polarization tomography provides for sensitive diagnostics of
array symmetry imperfections.

Finally, one may wonder whether depolarization in optics is
reversible or irreversible. In principle, depolarization is
reversible, as it is always due to some form of averaging over
spatial (or temporal) degrees of freedom\cite{Freund}; there is no
infinite bath that acts as an ``information sink''. In the present
experiment, ``repolarization'' would require an element that
modifies the array output polarization in a spatially-dependent
way; such an element could in principle be constructed based upon
a spatial light modulator\cite{Eriksen}, provided that it has a
sufficient number of degrees of freedom (pixels) available. So, in
the end, the difference between reversibility and irreversibility
is not absolute but gradual; it depends on the number of degrees
of freedom that can be managed in a practical case (The same
statement holds of course in statistical mechanics).

In conclusion, we have demonstrated surprising consequences of SP
propagation for the polarization behavior of nanohole arrays. The
non-locality of the array response forms an essential ingredient
of the physics of these intriguing devices.

\section*{Acknowledgements} This work has been supported by the
Stichting voor Fundamenteel Onderzoek der Materie (FOM); partial
support is due to the European Union under the IST-ATESIT
contract.

\end{document}